\documentclass[11pt,twoside]{article}

\usepackage{asp2006}
\usepackage{graphicx}

\begin{document}
\title{The micro-Jy Radio Source Population: the VLA-CDFS View}
\author{P. Padovani\altaffilmark{1}, V. Mainieri\altaffilmark{1,2},
P. Tozzi\altaffilmark{3}, K. I. Kellermann\altaffilmark{4},
E. B. Fomalont\altaffilmark{4}, N. Miller\altaffilmark{4,5,6},
P. Rosati\altaffilmark{1}, P. Shaver\altaffilmark{1}}

\altaffiltext{1}{European Southern Observatory, Karl-Schwarzschild-Str. 2,
D-85748 Garching bei M\"unchen, Germany}

\altaffiltext{2}{Max-Planck-Institut f\"ur Extraterrestrische Physik,
Giessenbachstr., D-85748, Garching bei M\"unchen, Germany}

\altaffiltext{3}{INAF, Osservatorio Astronomico di Trieste, Via G. B. Tiepolo 
11, I-34131, Trieste, Italy}

\altaffiltext{4}{National Radio Astronomy Observatory, 520 Edgemont Road, 
 Charlottesville, VA 22903-2475, USA}

\altaffiltext{5}{Department of Physics and Astronomy, Johns Hopkins
University, 3400 North Charles Street, Baltimore, MD 21218, USA}

\altaffiltext{6}{NRAO Jansky Fellow}

\begin{abstract}
We analyse the 267 radio sources from our deep (flux limit of 42 $\mu$Jy at
the field center at 1.4 GHz) Chandra Deep Field South 1.4 and 5 GHz VLA
survey. The radio population is studied by using a wealth of
multi-wavelength information, including morphology and spectral types, in
the radio, optical, and X-ray bands. The availability of redshifts for
$\sim 70\%$ of our sources allows us to derive reliable luminosity
estimates for the majority of the objects.  Contrary to some previous
results, we find that star-forming galaxies make up only a minority
($\approx 1/3$) of sub-mJy sources, the bulk of which are faint radio
galaxies, mostly of the Fanaroff-Riley I type.
\end{abstract}

\section{The VLA-CDFS Survey}   

The extragalactic radio source population ranges from normal galaxies with
luminosities $\sim 10^{20}$ W/Hz to galaxies whose radio emission is as
much as $10^{4 - 7}$ times greater owing to regions of massive star
formation or to an active galactic nucleus (AGN). The population of radio
sources in the sky with flux densities $> 1$ mJy is dominated by AGN driven
emission generated from the gravitational potential associated with a
supermassive black-hole in the nucleus. For these sources, the observed
radio emission includes the classical extended jet and double lobe radio
source as well as compact radio components more directly associated with
the energy generation and collimation near the central engine of the
AGN. Below 1 mJy there is an increasing contribution to the radio source
population from synchrotron emission resulting from relativistic plasma
ejected from supernovae associated with massive star formation in galaxies
or groups of galaxies, often associated with mergers or interactions
\citep[e.g.,][]{pad:win95,pad:ric98,pad:fom02}. However, the mix of star
formation and AGN related radio emission and the dependence on epoch is not
well determined, although the commonly accepted view ascribes to
star-forming galaxies a major role at these faint fluxes.

The Chandra Deep Field South (CDFS) area is part of the Great Observatories
Origins Deep Survey (GOODS) and as such is one of the most intensely
studied region of the sky. High sensitivity X-ray observations are
available from Chandra \citep{pad:gia02,pad:leh05}. The GOODS and Galaxy
Evolution from Morphology and Spectral energy distributions (GEMS)
multiband imaging programs using the HST Advanced Camera for Surveys (ACS)
give sensitive high resolution optical images
\citep{pad:gia04,pad:rix04}. Ground based imaging and spectroscopy are
available from the ESO 2.2m and 8m telescopes, and infrared observations
from the Spitzer space Telescope.

We observed the CDFS with the NRAO Very Large Array (VLA) for 50 hours at
1.4 GHz mostly in the BnA configurations in 1999 and 2001, and for 32 hours
at 5 GHz mostly in the C and CnB configurations in 2001. The effective
angular resolution was $3.5''$ and the minimum root-mean-square (RMS) noise
as low as $8.5~\mu$Jy per beam at 1.4 GHz and $7~\mu$Jy per beam at 5 GHz.
A total of 267 radio sources were catalogued at 1.4 GHz, 198 of which are
in a complete sample with signal-to-noise ratio (SNR) greater than 5 and
located within 15$^{\prime}$ of the field center. The corresponding flux
density limit ranges from 42 $\mu$Jy at the field center to 125 $\mu$Jy
near the field edge. These deep radio observations complement the larger
area, but less sensitive lower resolution observations of the CDFS
discussed by \cite{pad:afo06}.

Our set of ancillary data is quite unique and allows us to shed new light
on the nature of the sub-mJy radio source population. These data include:
a) reliable optical/near-IR identifications for $\sim 96\%$ of the radio
sources; b) optical morphological classification for $\sim 54\%$ of the
sample; of these: 38\% are spiral or interacting, 30\% are elliptical or
lenticular, 16\% are compact, and 15\% are irregular or complex; c)
redshift information for 186 ($\sim 70\%$) of the objects: 85 spectroscopic
(32\%) and 101 photometric (38\%). The redshift distribution covers the
range 0.03 -- 3.7 and peaks at $z \sim 0.6 - 0.7$, with a sharp cut-off at
$z\ga1$. The mean redshift is $\langle z \rangle \sim 0.8$; d) X-ray
detections for 85 ($\sim 32\%$) of the objects, upper limits for all the
others.

We present here preliminary results on the sub-mJy source population of the
VLA-CDFS. More details on the VLA-CDFS can be found in a series of four
papers, which address the radio data (Kellermann et al. 2007, in
preparation), the optical and near IR counterparts to the observed radio
sources (Mainieri et al. 2007, in preparation), their X-ray spectral
properties (Tozzi et al. 2007, in preparation), and the source population
(Padovani et al. 2007, in preparation). Throughout this paper spectral
indices are written $S_{\nu} \propto \nu^{-\alpha}$ and the values $H_0 =
70$ km s$^{-1}$ Mpc$^{-1}$, $\Omega_{\rm M} = 0.3$, and $\Omega_{\rm
\Lambda} = 0.7$ have been used.

\section{The sub-mJy number counts}

\begin{figure}[!ht]
\includegraphics[width=0.75\textwidth]{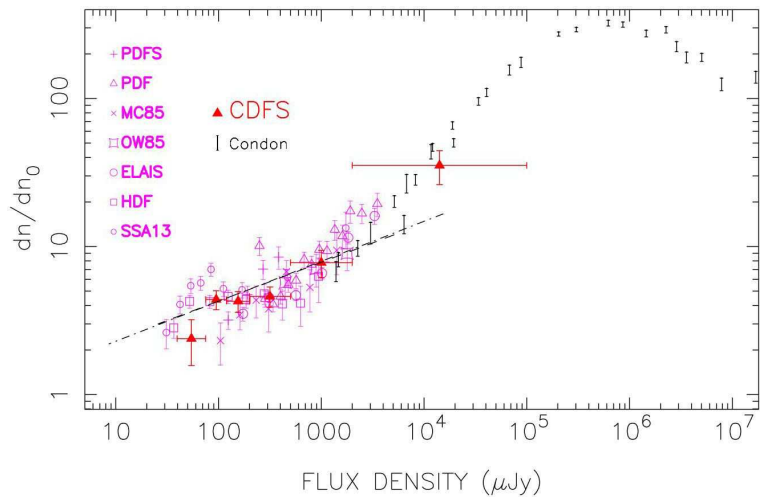}
\caption{The number counts at 1.4 GHz normalized to a Euclidean value. Our
data points are shown as triangles. The dashed line, which is the fit to
the CDFS counts, is a reasonable fit to the average counts from all surveys
at the lowest flux densities.}\label{pad:fig1}
 \end{figure}

Fig. \ref{pad:fig1} shows the counts of sources at 1.4 GHz over the entire
range of observed flux densities. The highest density points are based on a
compilation from Condon (2006, p.c.). Various studies over many decades
have shown that there are three regions of the counts: a) $>1$ Jy, where
the counts are rising with respect to Euclidean counts, due to the strong
evolution of luminous quasars and radio galaxies; b) 3 mJy -- 1 Jy, still
dominated by quasars and radio galaxies, and characterized by a drop-off
predominantly caused by the redshift effects and the cutoff above $z=3$ in
the density of these luminous radio sources; c) $< 3$ mJy, where the counts
flatten closer to Euclidean, well above the extrapolation from higher flux
densities, and the field to field scatter appears to increase below about
400 $\mu$Jy. This flattening of the slope of the counts, now observed in
the flux density range $35 - 300 \mu$Jy, is produced by a population of
objects, which is now dominant in the counts. The characterization of this
population is the major goal in the study of weak radio sources and of our
project.

\section{The sub-mJy population}

To shed light on the sub-mJy population, we start by evaluating a measure
of the radio-to-optical flux ratio, $R = \log (S_{\rm 1.4GHz}/S_{\rm
V})$. About $40\%$ of our sources have k-corrected $R$\footnote{$R$ values
have been k-corrected using templates typical for that class or, in the
case of sources without morphological classification, assuming a template
which was the average between ellipticals and spirals} values above the
star-forming regime, $R \la 1.4$, with some of our sources reaching
extremely high values of $R$, equivalent to radio-to-optical ratios $\ga
10^4 - 10^5$. It has been suggested that the radio-to-optical flux ratio
cannot be used to discriminate between AGN and star-forming galaxies
\citep[e.g.,][]{pad:afo05,pad:bar06}. We find this to be the case at low
($R \la 1.4$) values, where there is a large overlap between the two
classes. However, above this value practically only AGN are present amongst
the classified sources. Note that a sample with a brighter magnitude cutoff
would get very different results. For example, for $V \la 22$, only $\sim
10\%$ of the sources would have $R$ values beyond the star-forming
regime. This is an important point, which might explain some previous
results \citep[see also][]{pad:gru99}.

\begin{figure}[!ht]
\includegraphics[width=0.75\textwidth]{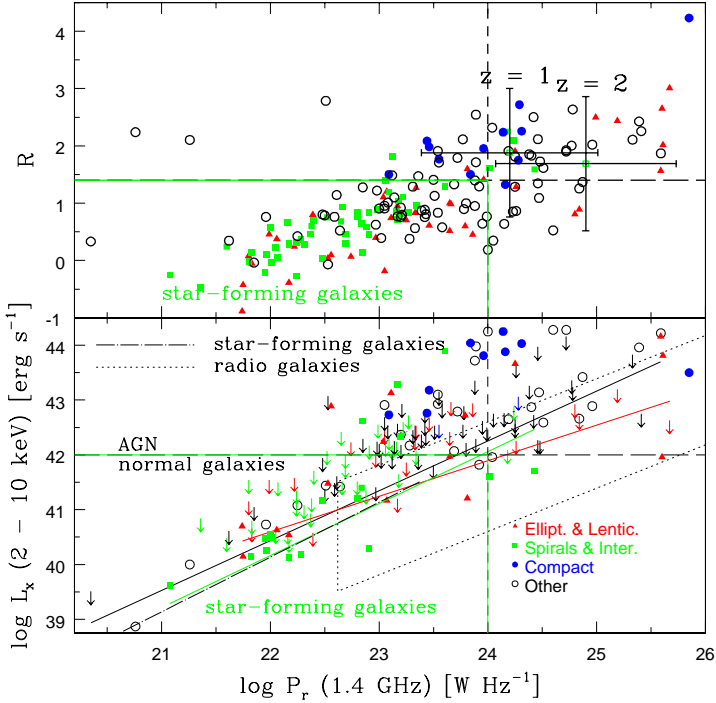}
\caption{{\it Top:} $R = \log (S_{\rm 1.4GHz}/S_{\rm V})$ vs. the 1.4 GHz
radio power for our sources.  The horizontal (long-dashed) line at $R =
1.4$ and the vertical (short-dashed) line at $\log P_{\rm r} = 24$ indicate
the maximum value for star-forming galaxies and the approximate dividing
lines between radio-loud and radio-quiet AGN. The mean values (and
$1\sigma$) for the sources without redshift information assuming $z=1$ and
$z=2$ are also shown. {\it Bottom:} The $2 - 10$ keV X-ray power vs. the
1.4 GHz radio power for sources with redshift and X-ray information. The
horizontal (long-dashed) line at $L_{\rm x} = 10^{42}$ erg/s indicates the
dividing line between AGN and normal galaxies. The dot-dashed line is the
$L_{\rm x} - P_{\rm r}$ relationship for nearby star-forming galaxies
\citep{pad:ra03}, converted to our cosmology, while the dotted lines denote the
locus of radio galaxies.  Arrows represent X-ray upper limits. Solid lines
are the best fits for the various classes taking upper limits into
account. \label{pad:fig2}}
 \end{figure}

Fig. \ref{pad:fig2}, top, which plots $R$ versus the 1.4 GHz radio power
for our sample, shows that the majority (92\%) of spiral and interacting
galaxies fall in the region typical of star-forming galaxies, i.e., $P_{\rm
r} \la 10^{24}$ W/Hz; most of them (90\%) have also $R < 1.4$. Ellipticals
and lenticulars span a very large range of radio powers, consistent with
that of low-luminosity (e.g., Fanaroff-Riley type I) radio-galaxies, while
all compact sources are consistent with being radio-loud AGN. Therefore,
despite the substantial overlap at low powers between ellipticals and
spirals, the various types have quite distinct radio powers. Namely,
$\langle \log P_{\rm r} \rangle = 22.65\pm0.10$, $23.46\pm0.19$,
$24.03\pm0.20$, and $23.59\pm0.11$ for spirals and interacting, ellipticals
and lenticulars, compact, and "other" sources (i.e., without morphological
classification) respectively. The latter have the largest $R$ values and a
distribution of radio powers which is significantly different ($P >
99.99\%$) from that of spiral and interacting galaxies, but similar to that
of ellipticals. Furthermore, the objects without redshifts have quite faint
optical magnitudes and are therefore very likely characterized by higher
than average redshifts. For estimated $z=1$ or $z=2$, the bulk of these
sources ends up in the region typical of radio galaxies.

Under the assumptions that: a) star-forming galaxies are identified with
spirals and interacting galaxies; b) star-forming galaxies alone occupy the
bottom-left quadrant of Fig. \ref{pad:fig2}, top, their fraction would be
$62^{+8}_{-7}\%$ of the objects with redshift information. However, only
$\sim 40\%$ of these sources have a definite spiral morphology, and their
maximum fraction (excluding ellipticals and lenticulars) is bound to be
$\la 3/4$. Moreover, as most of the objects without redshift information
have relatively large $R$ values and therefore are unlikely star-forming
galaxies, the total fraction of such objects in the sample is $< 45\%$.

Fig. \ref{pad:fig2}, bottom, shows $L_{\rm x}$ (2 -- 10 keV) vs. $P_{\rm
r}$ for the sources with redshift. Using ASURV \citep{pad:la92} to take into
account the upper limits on $L_{\rm x}$, we find a correlation between the
two powers for the whole sample ($L_{\rm x} \propto P_{\rm r}^{0.92\pm
0.08}$, $P > 99.99\%$). Moreover: a) the large majority (80\%) of spiral
and interacting galaxies have X-ray detections consistent with powers
typical of galaxies and not of AGN ($\la 10^{42}$ erg/s); the best fit
correlation using ASURV overlaps almost exactly with the X-ray-radio power
relationship for nearby star-forming galaxies (dot-dashed line); b) the
opposite is true for compact sources, which all have X-ray powers $ >
10^{42}$ erg/s and $\ga 30$ times larger than the other classes, at a given
radio power; this is consistent with the idea that these sources are mostly
radio-loud AGN (see also Fig. \ref{pad:fig2}, top); c) ellipticals and
lenticulars span a very large range of X-ray powers (similarly to what
observed in the radio band), consistent with that of radio-galaxies; again,
there is substantial overlap at low X-ray powers between ellipticals and
spirals, with at least 55\% of the former having $L_{\rm x} < 10^{42}$
erg/s; d) the "other" sources cover a very large range in radio and X-ray
powers, with the best fit correlation mostly into the radio-galaxies
area. Finally, as was the case for the radio powers, the various types have
quite distinct X-ray powers. Namely, $\langle \log L_{\rm x} \rangle =
40.65\pm0.15$, $41.34\pm0.21$, $43.40\pm0.18$, and $41.25\pm0.30$ for
spirals and interacting, ellipticals and lenticulars, compact, and "other"
sources respectively. The X-ray power distribution for the sources without
morphological classification is different ($P > 97\%$) from that of spiral
galaxies but similar to that of ellipticals.
 
Similarly to Fig. \ref{pad:fig2}, top, we can define the bottom-left
quadrant as that of star-forming galaxies. If star-forming galaxies alone
occupy this quadrant, their fraction would be $38^{+6}_{-5}\%$ of the
objects with redshift and X-ray information.  However, only about half of
these sources have a definite spiral morphology, and their maximum fraction
(excluding ellipticals and lenticulars) is bound to be $\la 3/4$. On the
other hand, some of the sources with upper limits above $L_{\rm x} \sim
10^{42}$ erg/s might have X-ray powers below that value. Therefore, the
total fraction of star-forming galaxies in the sub-sample with redshift
information is likely to be $\approx 30\%$.

\begin{figure}[!ht]
\includegraphics[width=0.49\textwidth]{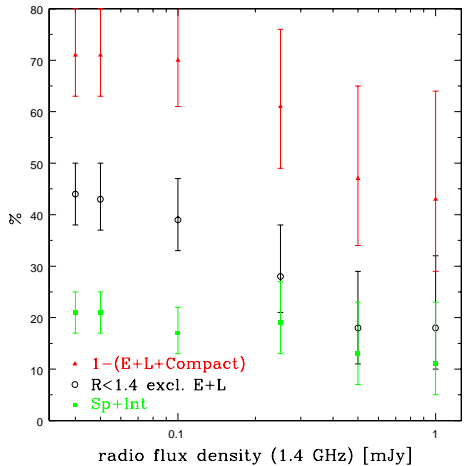}
\caption{The percentages of various morphological types as a function of
radio flux density for the complete sample.}\label{pad:fig3}
\end{figure}

Fig. \ref{pad:fig3} shows the percentage of various morphological types as
a function of radio flux density for the complete sample. The fraction of
spirals and interacting galaxies increases from $\sim 10\%$ at 1 mJy to
$\sim 20\%$ at the survey limit, while that of ellipticals, lenticulars,
and compact sources (i.e., the AGN, shown in the figure as $100-\%$) {\it
decreases} from $\sim 60\%$ to $\sim 30\%$. Hence, sub-mJy star-forming
galaxies down to $\sim 40 \mu$Jy can make up {\it at most} $70\%$ of the
total.  However, given the evidence presented above, a stronger upper limit
is provided by the fraction of sources having $R < 1.4$ excluding the
sources with elliptical and lenticular morphology, that is $\la
45\%$. Therefore, star-forming galaxies make up only a minority, between
$\sim 20$ and $45\%$ ($\approx 1/3$), of sub-mJy sources, the bulk of which
are faint radio galaxies, mostly at relatively low powers and hence of the
Fanaroff-Riley I type.

The VLA is operated by NRAO, a facility of the NSF under cooperative agreement 
with Associated Universities, Inc.



\end{document}